\documentclass[amsmath,amssymb,aps,pre,twocolumn,superscriptaddress]{revtex4-1}

\usepackage{graphicx}
\usepackage{dcolumn}
\usepackage{bm}
\usepackage{hyperref}
\usepackage{changes} 
\usepackage[utf8]{inputenc}
\usepackage{fancyhdr}
\definechangesauthor[color=purple, name={Paul Richter}]{P}

\newcommand*{\addheight}[2][.5ex]{%
  \raisebox{0pt}[\dimexpr\height+(#1)\relax]{#2}%
}

\newcommand\size{0.9}
\begin{document}
\title{Aging and equilibration in bistable contagion dynamics}

\author{Paul Richter}
\affiliation{Institute for Theoretical Physics, ETH Zurich, CH-8093 Zurich , Switzerland}
\email{richterp@ethz.ch}
\author{Malte Henkel}
\affiliation{Laboratoire de Physique et Chimie Th\'eoriques (CNRS UMR 7019), Universit\'e de Lorraine Nancy, B.P. 70239, F - 54506 Vand\oe uvre l\`es Nancy Cedex, France}
\email{malte.henkel@univ-lorraine.fr}
\affiliation{Centro de F\'isica T\'eorica e Computacional, Universidade de Lisboa, P - 1749-016 Lisboa, Portugal}
\affiliation{MPIPKS, N\"othnitzer Stra\ss e 38, D - 01187 Dresden, Germany}
\author{Lucas B\"ottcher}
\affiliation{Computational Medicine, UCLA, Los Angeles CA 90024, United States of America}
\affiliation{Institute for Theoretical Physics, ETH Zurich, CH-8093 Zurich , Switzerland}
\affiliation{Center of Economic Research, ETH Zurich, CH-8092, Zurich, Switzerland}
\email{lucasb@ethz.ch}

\date{\today}
\begin{abstract}
We analyze the late-time relaxation dynamics for a general contagion model. In this model, nodes are either active or failed. Active nodes can fail either ``spontaneously'' at any time or ``externally'' if their neighborhoods are sufficiently damaged.
Failed nodes may always recover spontaneously. At late times, the breaking of time-translation-invariance is a
necessary condition for physical aging. We observe that time-translational invariance is lost
for initial conditions that lie between the basins of attraction of the model's two stable stationary
states. Based on corresponding mean-field predictions, we characterize the observed model behavior
in terms of a phase diagram spanned by the fractions of spontaneously and externally failed nodes.
For the square lattice, the phases in which the dynamics approaches one of the two stable stationary
states are not linearly separable due to spatial correlation effects. Our results provide new insights
into aging and relaxation phenomena that are observable in a model of social contagion processes.
\end{abstract}
\maketitle

\section{Introduction}
The study of dynamical processes in complex systems is relevant in various contexts and contributed to a better understanding of the spreading of epidemics~\cite{keeling-rohani2008,boettcher14,satorras14,boettcher16, dehning2020inferring, PhysRevE.82.051921}, opinions~\cite{flache2017models}, innovations~\cite{rogers2003diffusion}, and other contagious phenomena. Seemingly different models that have been developed to describe the aforementioned processes share various (universal) properties. For some models, it is possible to observe such universal features in their relaxation dynamics. This behavior is also known as \emph{physical aging} and defined by (i) non-exponential, slow relaxation, (ii) breaking of time-translation invariance, and (iii) dynamical scaling~\cite{henkel08}. Examples of systems with a non-equilibrium steady state that exhibit aging include directed percolation~\cite{enss2004ageing,ramasco2004ageing,bottcher2018dynamical}, population dynamics~\cite{ageing2}, and gel-forming polymers~\cite{gel1}. 

Identifying universal dynamical features in non-equilibrium systems may be useful to make predictions about their long-time behavior and, in the case of contagious processes, develop control and intervention protocols. Here we focus on the aging characteristics of a general contagion model that captures a variety of simple (i.e., single contact~\cite{tome10,carletti2020covid,willis2020insights,dehning2020inferring}) and complex (i.e., multiple contact~\cite{boettcher162,boettcher171}) contagion dynamics in terms of spontaneous and externally-induced infection/failure processes. Models of simple contagions are common tools to describe the spread of epidemics~\cite{keeling-rohani2008}. In contrast to simple contagions where one infected individual is able to infect others, complex (or social) contagions require contact with multiple ``infected'' individuals~\cite{granovetter78,macy07}. Examples of complex contagions include the diffusion of innovations~\cite{coleman57,rogers2003diffusion}, political mobilization~\cite{chwe99}, viral marketing~\cite{leskovec07}, and coordination games~\cite{easley2010}.

Previous studies~\cite{enss2004ageing,ramasco2004ageing,bottcher2018dynamical} investigated the aging properties of simple contagions whose phase space consists of an absorbing and a fluctuating phase with unique stationary states. Within these two phases, relaxation is exponential whereas slow (algebraic) relaxation can be observed at the critical point where the two phases merge~\cite{henkel08}. For complex contagions, the phase space cannot be described by two phases with unique stationary states. Instead, it is characterized by a bistable regime~\cite{majdandzic14,boettcher162,boettcher171} that gives rise to relaxation towards one of two stable stationary states.

In this work, we study relaxation and aging kinetics of complex contagions within their bistable region. In Sec.~\ref{sec:model}, we introduce the general contagion model and provide an overview of corresponding mean-field results and concepts from the study of aging. In Sec.~\ref{sec:relax_dynamics}, we numerically determine a phase portrait of bistable contagion dynamics on a square lattice and study the influence of different initial densities of failed nodes on the relaxation characteristics. In Sec.~\ref{sec:phasediagram}, to provide more insights into the relaxation properties, we map the observed dynamics to a phase diagram spanned by the fractions of spontaneously and externally-failed nodes. We find that the dynamics is initially driven by spontaneous transitions and rapidly approaches a line along which externally-induced transitions slowly drive the system to one of the two stable stationary states. We conclude our study and discuss our results in Sec.~\ref{sec:conclusion}.
\section{Model and Methods}
\label{sec:model}
\begin{figure}
\centering
\includegraphics[width = 0.3\textwidth]{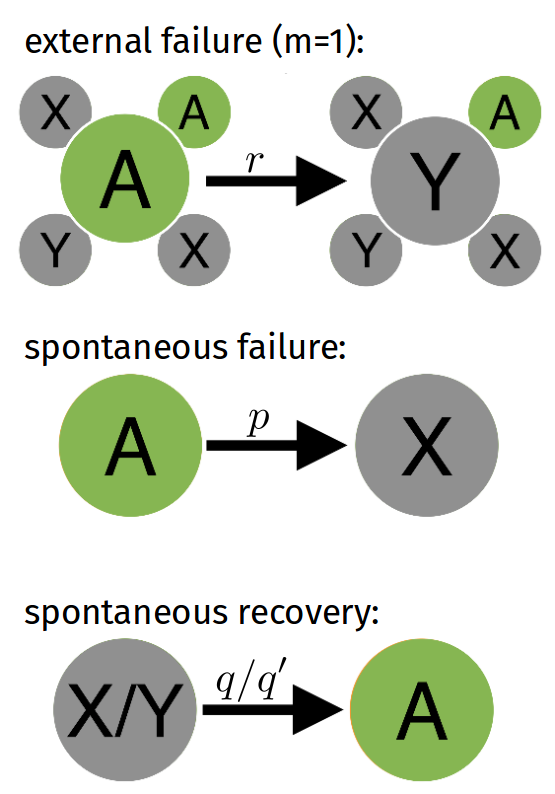}
\caption{\textbf{Schematic of model dynamics}. Nodes are arranged in a square lattice. Active nodes ($A$) can fail if their neighborhoods are sufficiently damaged (external failure) with rate $r$ or spontaneously (spontaneous failure) with rate $p$. A node fails externally if less than or equal to $m$ of its neighbors are active. The recovery rate of spontaneously-failed nodes is $q$ and that of externally-failed nodes is $q'$. We use $X$ and $Y$ to indicate that nodes failed spontaneously and externally, respectively.}
\label{fig:scheme}
\end{figure}
We consider a general contagion model on a network whose $N$ nodes are either active ($A$) or failed ($X$ or $Y$)~\cite{boettcher162,boettcher171}. Active nodes can fail ``spontaneously'' with rate $p$ or ``externally'' with rate $r$ if their neighborhoods are sufficiently damaged. We use $X$ and $Y$ to denote the state of nodes that failed spontaneously and externally, respectively. External failure occurs if the number of active nearest neighbors of a node is smaller than or equal to the threshold $m$. Failed nodes in states $X$ and $Y$ recover with rates $q$ and $q'$, respectively. We illustrate the described failure and recovery processes in Fig.~\ref{fig:scheme}.

To formulate the corresponding mean-field rate equations, let $n(t)\in [0,1]$ be the fraction of active nodes, which is one minus the sum of the fractions of nodes $u(t)$ and $v(t)$ that failed spontaneously and externally (i.e., $n(t)=1-u(t)-v(t)$). We use $n_{\text{st}}$ to denote the total fraction of active nodes in the stationary state. For the derivation of the mean-field rate equations, we assume perfect mixing and first focus on the spontaneous-failure dynamics
\begin{equation}
\frac{\mathrm{d} u(t)}{\mathrm{d} t}= p n(t)- q u(t)\,,
\label{eq:gen_cont__spon}
\end{equation}
where the first term accounts for spontaneous failure with rate $p$ and the second term accounts for spontaneous recovery with rate $q$. 

Next, we use the term \emph{critically-damaged neighborhood} (CDN) to refer to a neighborhood where the number of active neighbors is smaller than or equal to $m$. The probability that a node of degree $k$ is located in a CDN is $E_k=\sum_{j=0}^m  \binom {k} {k-j} (1-n)^{k-j} n^j$.  Therefore, the time evolution of externally-failed nodes is given by
\begin{equation}
\frac{\mathrm{d} v(t)}{\mathrm{d} t}= r \sum_k f_k E_k n(t)- q' v(t)\,,
\label{eq:gen_cont__ind}
\end{equation}
where $f_k$ is the degree distribution of the underlying network. The first term of Eq.~\eqref{eq:gen_cont__ind} describes that active nodes become inactive with rate $r$ if their neighborhood contains a sufficient number of inactive nodes and the second term accounts for spontaneous recovery with rate $q'$. This model is close in spirit to the well-studied simple SIR model~\cite{tome10} which despite its simplicity has been used recently to study the propagation of pandemics~\cite{carletti2020covid,willis2020insights}. 

The outlined mean-field theory provides a reference point to mathematically characterize the stationary and relaxation properties of complex contagions. In the following sections, we will mainly focus on complex contagion dynamics on the square lattice, for which mean-field results can only qualitatively capture the observed behavior (see Ref.~\cite{boettcher162} for a detailed comparison of mean-field predictions for complex contagions and corresponding simulation result on different random and spatially-embedded networks). Most of our subsequent analyses will therefore be based on simulation results.

For regular networks with degree $k$, a hysteresis region exists for $m < k-1$~\cite{boettcher162,boettcher171}. Within this region, there are three stationary states with densities of active nodes $n_{\mathrm{st}}^1$, $n_{\mathrm{st}}^2$, and $n_{\mathrm{st}}^3$. The stable stationary states have densities $n_{\mathrm{st}}^1$ and $n_{\mathrm{st}}^3$ and the density of the unstable state is $n_{\mathrm{st}}^2$. We shall outline in Sec.~\ref{sec:phasediagram} that the unstable stationary state is best characterized by the corresponding densities of externally and spontaneously failed nodes $(u_{\rm st}^2,v_{\rm st}^2)$.

In addition to the outlined complex contagion dynamics, Eq.~\eqref{eq:gen_cont__ind} can be also used to model purely spontaneous dynamics ($m=k$) and simple contagions ($m=k-1$). If $m=k$, external failures even occur if all neighbors of a certain node are active. Thus, for $m=k$ all failure and recovery processes are spontaneous. If $m=k-1$, active nodes may externally fail if at least one of its neighbors is in a failed state ($X$ or $Y$). This process is thus connected to simple contagions~\cite{keeling-rohani2008,boettcher171,tome10,dehning2020inferring,carletti2020covid,tome2003role} where individuals may become infected if they were in contact with at least one infected person. One formulation of simple contagions is the contact process, which is a common model of absorbing phase transitions in non-equilibrium statistical physics~\cite{henkel2008non}. In mathematical epidemiology~\cite{keeling-rohani2008}, simple contagion processes can be found in most epidemic models including the susceptible-infected-susceptible model and the susceptible-infected-recovered (SIR) model.

All numerical mean-field solutions are based on an explicit Euler forward integration scheme with time step $\Delta t = 0.01$. To simulate the described reactions on a square lattice of linear dimension $L$ and with $N=L\times L$ nodes, we use kinetic Monte-Carlo (i.e., Gillespie) methods~\cite{gillespie76,gillespie77}. At time $t$, each node $i\in \{1,2,\dots,N\}$ is either active or failed. In our simulations, we keep track spontaneously and externally failed nodes and indicate the failure of node $i$ at time $t$ by $n_i(t)=0$. Similarly, we indicate an active state of node $i$ at time $t$ by $n_i(t)=1$.

To compare the relaxation properties that result from initial conditions close to the stable and unstable states, we quantify the relaxation dynamics of the described contagion model in terms of the autocorrelation function
\begin{align}
\Gamma(t,s) &= \langle n_i(t) n_i(s)\rangle\label{eq:gamma}
\end{align}
and autocovariance function
\begin{align}
C(t, s) &= \langle n_i(t) n_i(s) \rangle - \langle n_i(t)\rangle  \langle n_i(s)\rangle\,, \label{eq:C}
\end{align}
where angular brackets $\langle \cdot \rangle$ denote an ensemble average. In systems that exhibit aging, both $\Gamma$ and $C$ do not depend on $t-s$ alone, but are also expected to obey the following scaling behavior~\cite{henkel08,bottcher2018dynamical}:
\begin{align}
\Gamma(t,s) &= s^{-b} f_{\Gamma}(t/s) \, \,, \,f_{\Gamma}(y) \sim y^{-\lambda_\Gamma/z}\,,\label{eq:gamma2} \\
C(t,s) &= s^{-b} f_{C}(t/s) \, \,, \,f_{C}(y) \sim y^{-\lambda_C/z}\,,\label{eq:C2}
\end{align}
for $t,s \gg \tau_{\text{micro}}$ and $t-s \gg \tau_{\text{micro}}$, where $\tau_{\text{micro}}$ is a microscopic reference time scale. Equivalently, one may also study the critical short-distance dynamics and compute the initial slip exponent $\Theta$, as was done in Refs.~\cite{tome10,argolo2012critical} for a simple contagion model. The exponent $z$ is the dynamical exponent and the autocorrelation exponents $\lambda_\Gamma$ and $\lambda_C$ are defined from the asymptotics for $y=t/s \gg 1$ of the corresponding scaling functions.

In the context of failure and opinion spreading in technical and social networks, the autocorrelation and autocovariance functions $\Gamma$ and $C$ are useful to characterize how such contagious phenomena progress over time~\cite{enss2004ageing,ramasco2004ageing,bottcher2018dynamical,henkel08}. The autocorrelation function $\Gamma(t,s)$ quantifies the probability of a local node failure/infection at time $t$ after a local node failure/infection a time $s$. In the definition of the autocovariance function $C(t,s)$, uncorrelated time evolutions $\langle n_i(t) \rangle$ and $\langle n_i(s) \rangle$ are removed from the autocorrelation data. Instead of accounting for local densities, one can also define global correlators which are based on the prevalence of failure, infections, and opinions in the whole population~\cite{bottcher2018dynamical}.
\section{Relaxation dynamics}
\label{sec:relax_dynamics}
\begin{figure*}
\centering
\includegraphics[scale = 1]{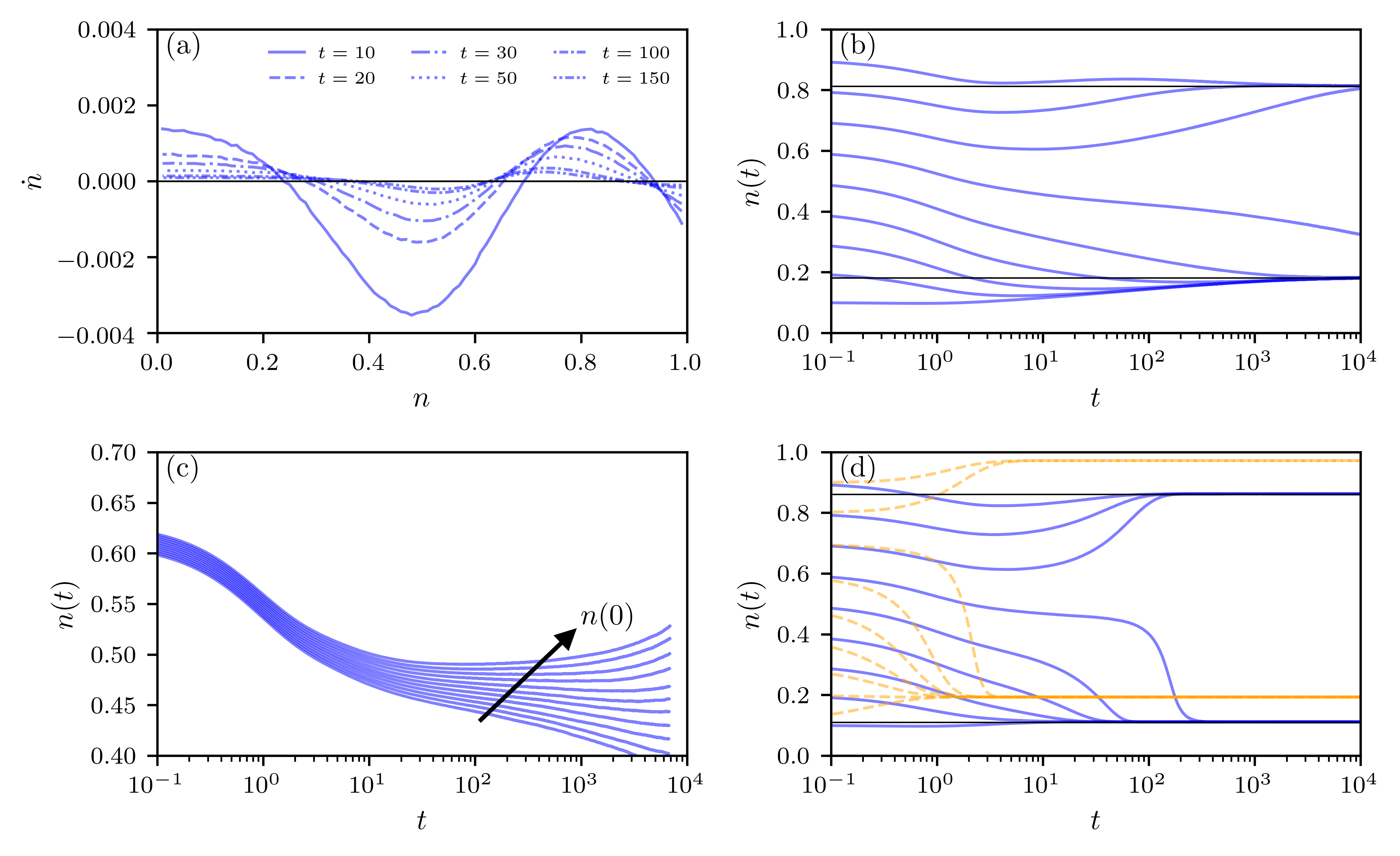}
\caption{\textbf{Phase portrait and relaxation dynamics}. (a) The numerically-determined phase portrait of the fraction of active nodes $n(t)$ in the general contagion model. Simulations were performed on a square lattice with $N=512\times512$ sites using $10^3$ samples for each data point. (b) The time evolution of $n(t)$ for different initial conditions on a square lattice with $N=1024\times 1024$ sites. We use $n(0) \in \{0.1, 0.2,\ldots, 0.9\}$ as initial densities of active nodes. (c) The relaxation dynamics of the general contagion model on a square lattice with $N=1024\times 1024$ nodes is shown. The chosen initial densities are close to $n^*(0) \approx 0.62$. For initial densities $n(0) > n^*(0)$, we observe that $n(t)$ moves towards the upper stationary state, while for $n(0) < n^*(0)$ the densities move towards the lower stationary state. We use $n(0) \in \{ 0.610, 0.612, 0.614, \ldots , 0.630 \}$ as initial densities. Since the time axis begins at $t=10^{-1}$ and not at $t=0$, the densities have already slightly decreased.
(d) The time evolution of $n(t)$ for different initial conditions on a random regular graph with $N=10^5$ nodes and degree $k = 4$ (blue solid lines) and corresponding mean-field results (orange dashed lines). The initial conditions are the same as in
panel (b), but the relaxation is significantly faster than on a square lattice. In all simulations, we set $m=1$, $r=0.95$, $p=1.0$, $q=1.0$, and $q'=0.1$. The initial density of spontaneously-failed nodes is zero and we use uniformly-distributed externally-failed nodes such that $v(0)=1-n(0)$.
}
\label{fig:graphs1}
\end{figure*}
With respect to simple models of infection, such as directed percolation~\cite{enss2004ageing,ramasco2004ageing,bottcher2018dynamical,henkel08} or the SIR model~\cite{tome10,carletti2020covid,dehning2020inferring} whose phases have each a {\em single} stationary state, a complex contagion as described by the general contagion model studied here exhibits a bistable regime for certain thresholds and failure and recovery rates~\cite{boettcher162,boettcher171}. One possible choice is to set $m=1$, $r=0.95$, $p=1.0$, $q=1.0$, and $q'=0.1$ to obtain bistable dynamics on a square lattice. We use these parameters for our simulations on the square lattice throughout the manuscript. Note that the extent of the bistable regime is rather small for the square lattice and other embedded networks, whereas it is substantially larger for random networks and the mean-field case~\cite{boettcher162,boettcher171}. Initially, we set the density of spontaneously-failed nodes to zero and use uniformly distributed externally-failed nodes. The initial density of active nodes is thus $n(0)=1-v(0)$. Since we are interested in the relaxation properties of the model dynamics within this bistable region, we first determine the initial density of active nodes $n^*(0)$ from which the system eventually relaxes into one of the two stable stationary states with densities $n_{\mathrm{st}}^1$ and $n_{\mathrm{st}}^3$. 
Mathematically, $n^*(0)$ is the separatrix in the initial density $n(0)$ such that for $n(0)>n^*(0)$, the system evolves toward the steady state with density $n_{st}^1$
    and for $n^*(0)<n(0)$, the system evolves towards the steady state with density $n_{st}^3$.
Note that $n^*(0)$ is not the same as the density of active nodes in the unstable density of states (see Sec.~\ref{sec:phasediagram}).

After initializing the dynamics, we numerically construct a phase portrait of $n(t)$ by approximating the slope $\dot{n}$ in terms of the finite-difference derivative $(n(t+\Delta t)-n(t))/\Delta t$ with $\Delta t=0.1 t$. Based on the behavior of $\dot{n}$ in Fig.~\ref{fig:graphs1}(a), we can identify all stationary states (i.e., the states for which the density of active nodes satisfies $\dot{n}=0$). Close to the initial density of active nodes $n^*(0)$, the slope $\dot{n}$ changes its sign from negative to positive values and we conclude that $n \approx 0.62=n^*(0)$. To determine the densities of the two stable stationary states, we perform simulations on a square lattice with $1024\times 1024$ nodes. In Fig.~\ref{fig:graphs1}(b), we show the corresponding time evolution of the density of active nodes $n(t)$ for different initial conditions. For the considered failure rates, we find that the densities of active nodes in the two stable stationary state are $n_{\text{st}}^1\approx 0.18$ and $n_{\text{st}}^3 \approx 0.81$. For a comparison between complex contagion relaxation dynamics on a square lattice with those on a random graph, we also simulated the propagation of a complex contagion within its bistable region on a random regular graph with $N=10^5$ nodes and degree $k=4$ (see Fig.~\ref{fig:graphs1}(d)). For sufficiently large degrees $k$, it has been shown in detail~\cite{boettcher162} that complex contagion on random regular graphs is indeed well-described by the corresponding mean-field theory. For a relatively small degree such as $k=4$, the qualitative features of the relaxation are similar. However, because of the smaller diameter of the random regular network under consideration, its relaxation is significantly faster than on a square lattice, as illustrated in Fig.~\ref{fig:graphs1}(d). The perfect-mixing assumption that underlies the formulation of the mean-field rate equations \eqref{eq:gen_cont__spon} and \eqref{eq:gen_cont__ind} entails mean-field relaxation dynamics (orange dashed lines in Fig.~\ref{fig:graphs1}(d))  which are even faster than the relaxation dynamics on the random regular graph with degree $k=4$.

The numerically-obtained phase portrait of the square lattice is qualitatively very similar to corresponding mean-field results~\cite{boettcher171}. However, in the following sections, we show that spatial correlation effects on the square lattice lead to subtle deviations of the initial relaxation dynamics from mean-field predictions. 
In Fig.~\ref{fig:inex}(c), we provide a more detailed picture of the relaxation dynamics in the vicinity of $n^*(0)$. If the initial fraction of active nodes is larger than $n^*(0) \approx 0.62$, the curves move towards the upper stable stationary states; for initial fractions below $n^*(0)$ they approach the lower point of stability. The closer the initial density is to $n^*(0) \approx 0.62$, the slower it approaches one of the stable stationary states. This behavior can also be understood in terms of the phase portrait in Fig.~\ref{fig:graphs1}(a). If the system is sufficiently close to $n^*(0)$, the slope $\dot{n}$ approaches zero.

The slow relaxation dynamics in systems that exhibit aging is mathematically often described by a power law $n(t) \sim t^{-\delta}$ and we find $\delta \approx 0$ for the equilibration behavior close to the unstable stationary state. We observe in Fig.~\ref{fig:graphs1}(b) that the fraction of active nodes may first decrease and then increase again for certain initial conditions. In the next section, we further investigate this behavior by illustrating the system's time evolution in terms of a phase diagram spanned by the fractions of spontaneously and externally-failed nodes.
\section{Phase diagram}
\label{sec:phasediagram}
\begin{figure*}
\centering 
\includegraphics[scale = 1]{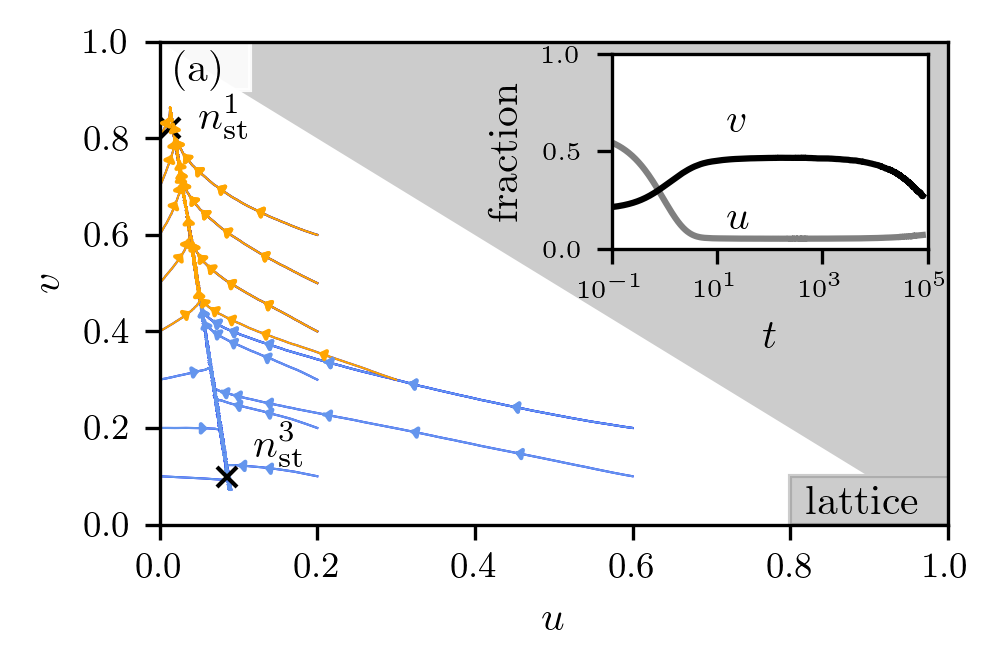}
\includegraphics[scale = 1]{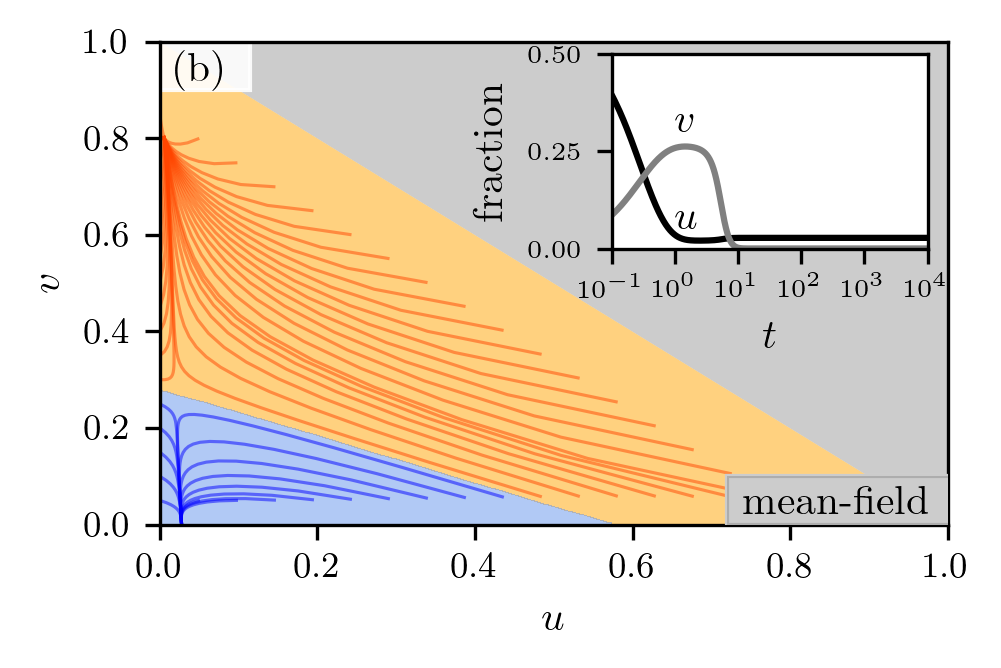}
\includegraphics[scale = 1]{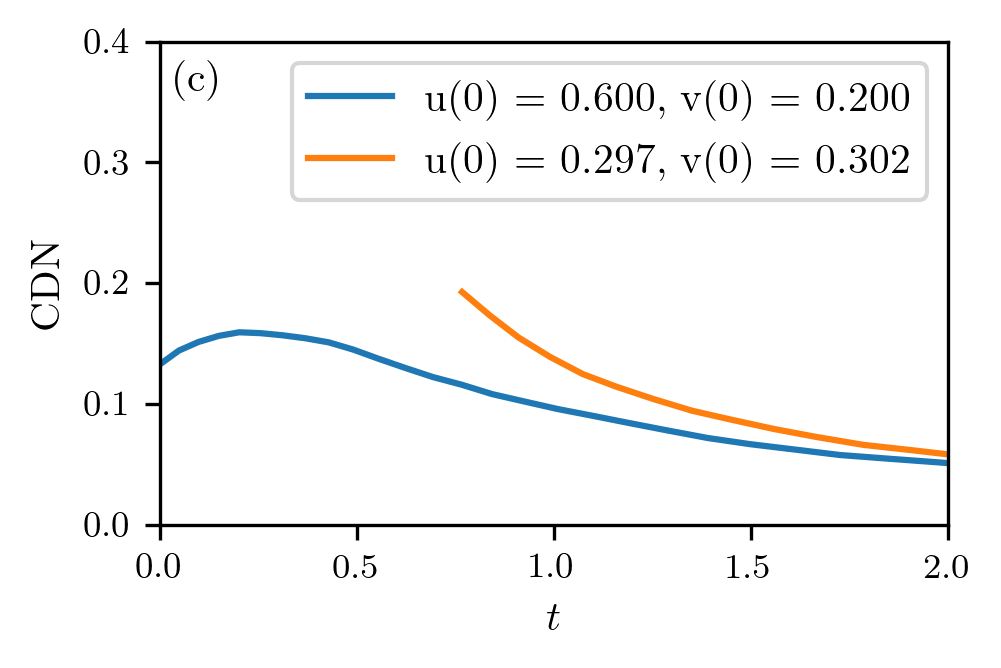} 
\includegraphics[scale = 1]{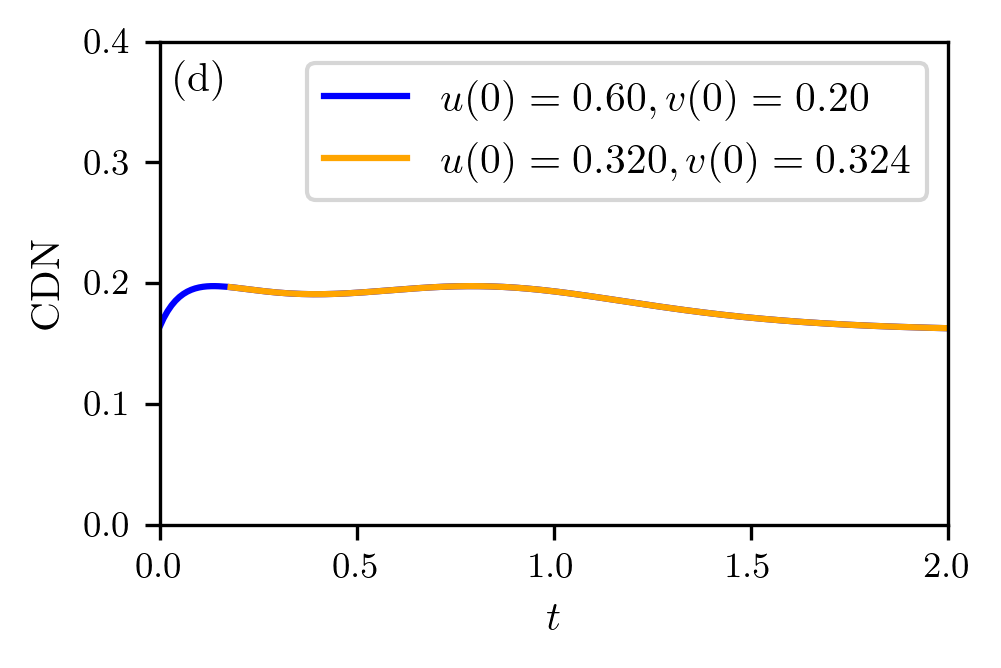}
\caption{
\textbf{Relaxation for different fractions of externally and internally failed nodes.} (a) The relaxation of the model dynamics in the $u-v$ plane for different initial conditions. All trajectories first approach an ``equilibration line'' (stage I) and then slowly relax towards one of the two stable stationary states that we indicate by black crosses (stage II). Blue and orange trajectories approach the stationary states with densities $n_{\rm st}^1$ and $n_{\rm st}^3$, respectively. Simulations were performed on a square lattice with $N=1024\times 1024$ nodes and for parameters $m=1$, $r=0.95$, $p=1.0$, $q=1.0$, and $q'=0.1$. (b) Corresponding mean-field trajectories (black solid lines) in the $u-v$ plane for different initial conditions and $m=1$, $r=5.0$, $p=0.1$, $q=3.5$, and $q'=1.0$. Within the orange (blue) region, all trajectories approach the stationary state with density $n_{\rm st}^1$ ($n_{\rm st}^3$). To obtain the mean-field trajectories, we numerically solve Eqs.~\eqref{eq:gen_cont__spon} and \eqref{eq:gen_cont__ind}. In both panels, grey-shaded regions correspond to densities $n>1$ and are excluded from our analysis. (c) The fractions of CDNs for the two trajectories that cross in (a). For initial conditions $(u(0)=0,v(0))=(0.6,0.2)$ (blue solid line), we observe that the proportions of CDNs are smaller than for $(u(0.77)=0,v(0.77))\approx(0.3,0.3)$. (d) If two mean-field trajectories share the same densities $u$ and $v$, they converge towards the same stationary state.}
 \label{fig:inex}
\end{figure*}
To better understand the initial relaxation dynamics, we study the evolution of the density of active nodes in the $u-v$ plane. In Fig.~\ref{fig:inex}(a), we show the evolution of the densities $u(t)$ and $v(t)$ for different initial conditions on a square lattice. We observe that the evolution of all trajectories can be divided into two stages I and II. During stage I, the evolution of a trajectory is mainly driven by spontaneous-failure dynamics. After some time, all trajectories approach an ``equilibration line'' and relax towards one of the two stable stationary states (stage II). The relaxation during stage I is much faster than the relaxation in stage II (see inset in Fig.~\ref{fig:inex} (a)). We find the following densities of spontaneously- and externally-failed nodes at the two stable stationary states
\begin{align}
\begin{split}
u^1_{\rm st} = 0.019(1)\quad\text{and}\quad v^1_{\rm st} = 0.803(1)\,, \\
u^3_{\rm st} = 0.087(1) \quad\text{and}\quad v^3_{\rm st} = 0.100(1)\,.
\end{split}
\end{align}
Note that the corresponding densities of active nodes $n^1_{\rm st}=1-u^1_{\rm st}-v^1_{\rm st}=0.178(1) $ and $n^3_{\rm st}=1-u^3_{\rm st}-v^3_{\rm st}= 0.813(1)$ are equal to the values that we reported in Sec.~\ref{sec:relax_dynamics}. Although the stable stationary states with densities $n^1_{\rm st}$ and $n^3_{\rm st}$ can be identified with the phase portrait and evolution plots of $n(t)$ (see Fig.~\ref{fig:graphs1}), the situation is more complex for the characterization of the unstable stationary state. 

For the simulations in Fig.~\ref{fig:graphs1}, we set $u(0)=0$ and use uniformly-distributed externally failed nodes. The actual unstable point lies at $(u^2_{\rm st},v^2_{\rm st})=(0.060(5),0.370(10))$ and has a density of active nodes $n^2_{\rm st}=0.570(15)$. The point $n^*(0)$ (i.e., $(u(0),v(0))=(0,0.38)$) of Sec.~\ref{sec:relax_dynamics} lies at the boundary that separates the phases in which they dynamics either approaches the upper or the lower stable stationary state.

On the square lattice, we find additional effects that make the dynamics even more complex. We observe in Fig.~\ref{fig:inex}(a) that the evolution of $u(t)$ and $v(t)$ is not fully determined by these two densities alone. For $(u,v)\approx (0.3,0.3)$, the blue trajectory starting at $(u(0),v(0))=(0.6,0.2)$ and ending at $(u^3_{\rm st},v^3_{\rm st})$ intersects with the orange trajectory, which converges towards $(u^1_{\rm st},v^1_{\rm st})=(0.019,0.803)$. Based on this result, we conclude that the evolution of $(u(t),v(t))$ on a square lattice depends on the initial densities $(u(0),v(0))$ even if two trajectories share the same densities at some time. In other words, it is not possible to describe the observed dynamics on a square lattice in terms of differential equations of $u(t)$ and $v(t)$. The described effect can be also understood in terms of the fractions of CDNs, as we show in Fig.~\ref{fig:inex}(c). Initially, we distribute all externally-failed nodes uniformly at random on the lattice. The corresponding initial density of CDNs is $\mathrm{CDN}(0)\approx 0.13$. At time $t \approx 0.77$, the aforementioned blue and orange trajectories intersect. At this time, the density of CDNs of the blue trajectory is $\mathrm{CDN}(0.77)\approx 0.1$, whereas the (initial) density of CDNs of the orange trajectory is $\mathrm{CDN}(0.77)\approx 0.2 $.

For a qualitative comparison, we solve the mean-field rate equations of $u(t)$ and $v(t)$ (see Eqs.~\eqref{eq:gen_cont__spon} and \eqref{eq:gen_cont__ind}) for parameters that lead to a bistable contagion dynamics. We show the corresponding $u-v$ phase space in Fig.~\ref{fig:inex} (b). Similar to the observation that we made for the square-lattice case, the inset in Fig.~\ref{fig:inex} (b) shows that the relaxation of the spontaneous-failure dynamics is faster than the induced-failure dynamics. This mean-field analysis also enables us to identify two distinct phases. All trajectories that originate within the blue region approach the stable stationary state with density $n^3_{\rm st}$. The remaining trajectories that originate in the orange region move towards the second stable stationary state density $n^1_{\rm st}$. Unlike in the square-lattice phase space, both phases can be linearly separated, because the mean-field system is fully determined by the densities $u(t)$ and $v(t)$ since the mean-field dynamics is described by deterministic differential equations (see Eqs.~\eqref{eq:gen_cont__spon} and \eqref{eq:gen_cont__ind}) with unique trajectories. Thus, if two trajectories share the same value of $u(t)$ and $v(t)$, they also have the same fractions of CDNs (see Fig.~\ref{fig:inex}(d)). 
\section{Correlation effects}
In Secs.~\ref{sec:relax_dynamics} and \ref{sec:phasediagram}, we outlined the influence of different initial conditions on the relaxation characteristics of the general contagion model. In this section, we study the properties of the autocorrelation function $C$ and autocovariance function $\Gamma$ (see Eqs.~\eqref{eq:gamma} and \eqref{eq:C}). First, we initialize the dynamics with $n(0) = 0.62 \approx n^*(0)$ and $n(0)=0.81 \approx n_{\mathrm{st}}^3$ on a square lattice with $512\times 512$ sites. The corresponding initial densities of spontaneously- and externally-failed nodes are $(u(0),v(0))=(0,0.38)$ and $(u(0),v(0))=(0,0.19)$. Next, we generate $10^3$ sample trajectories and compute the corresponding values of $C$ and $\Gamma$. 

\begin{figure*}[h]
\centering
\begin{tabular}{ccc}
\addheight{\includegraphics[scale = \size]{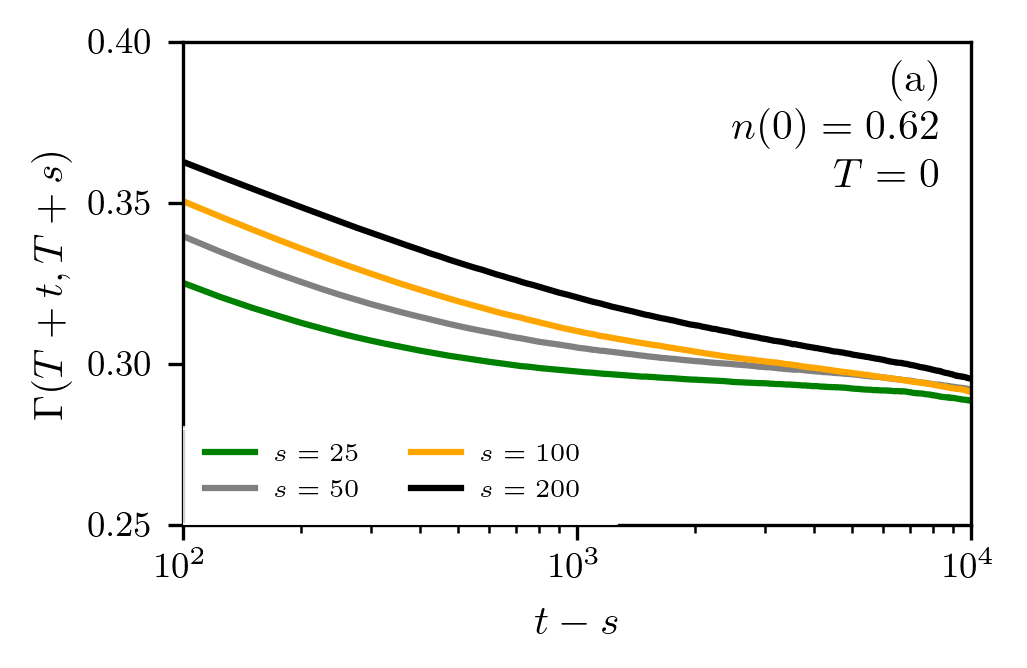}}&
\addheight{\includegraphics[scale = \size]{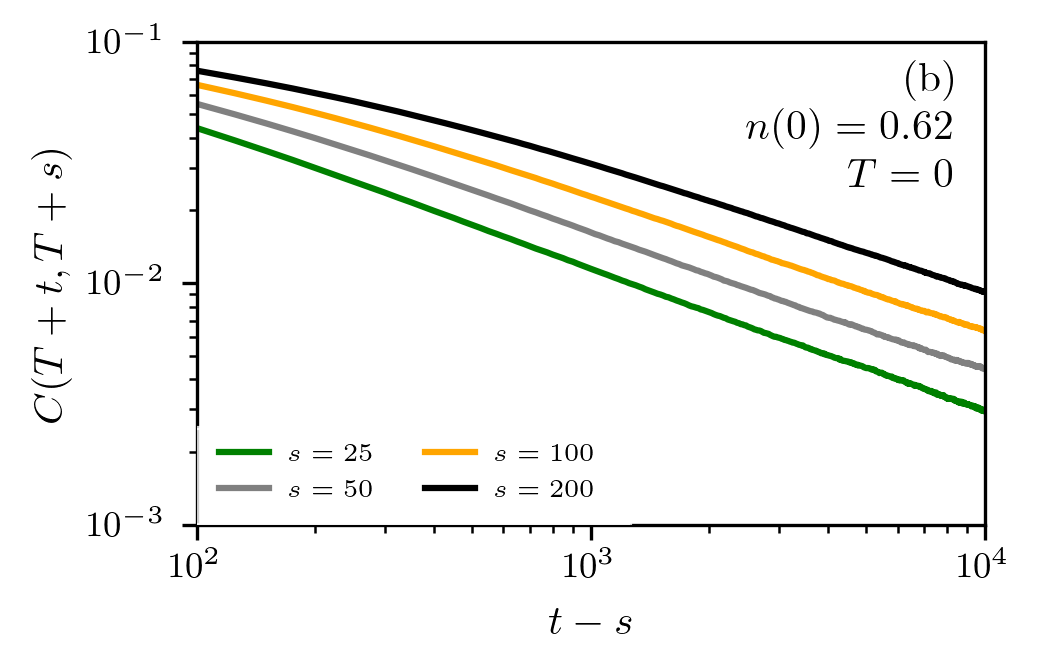}}\\
\addheight{\includegraphics[scale = \size]{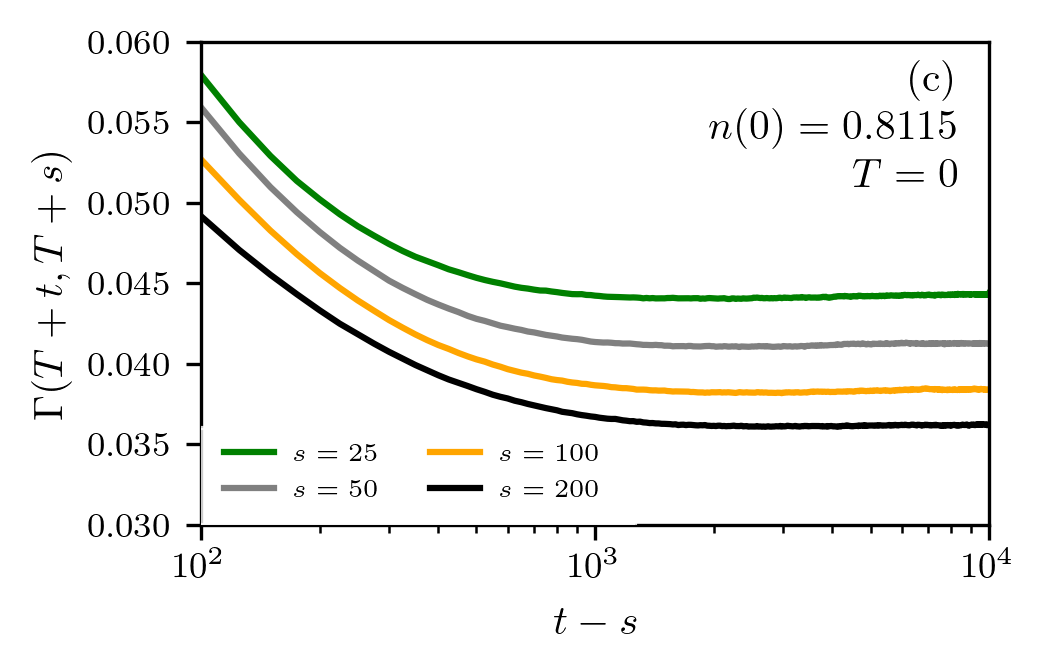}}&
\addheight{\includegraphics[scale = \size]{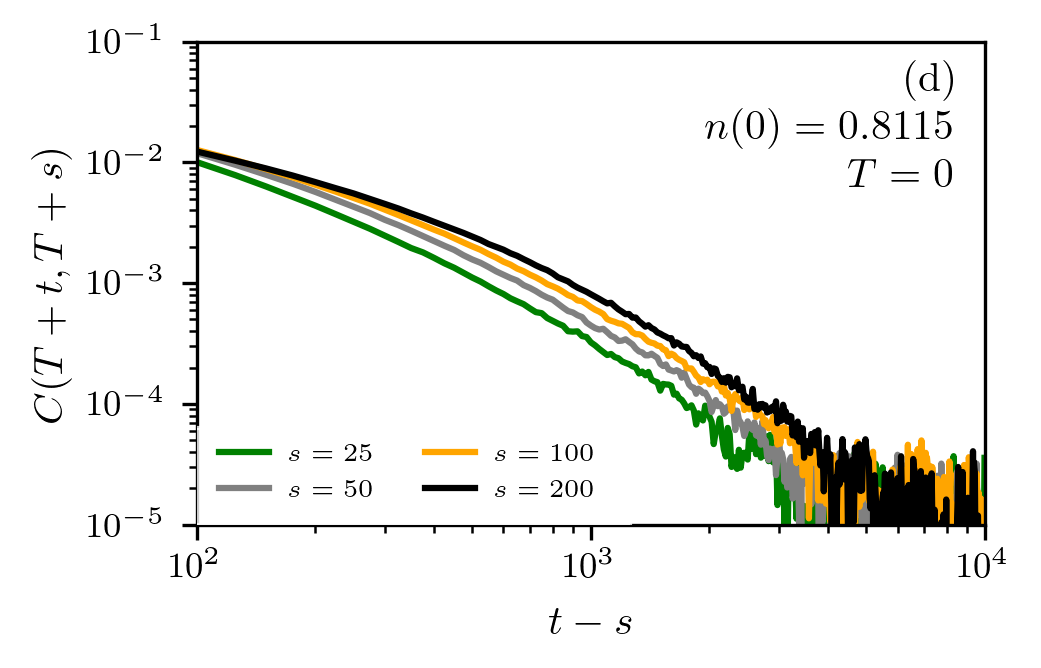}}\\
\addheight{\includegraphics[scale = \size]{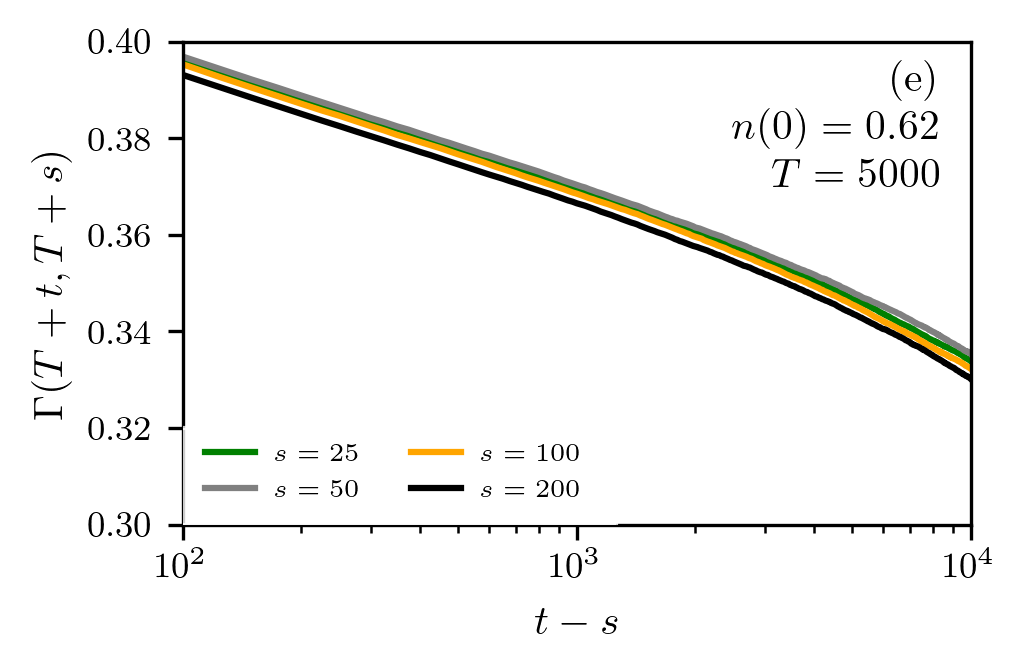}}&
\addheight{\includegraphics[scale = \size]{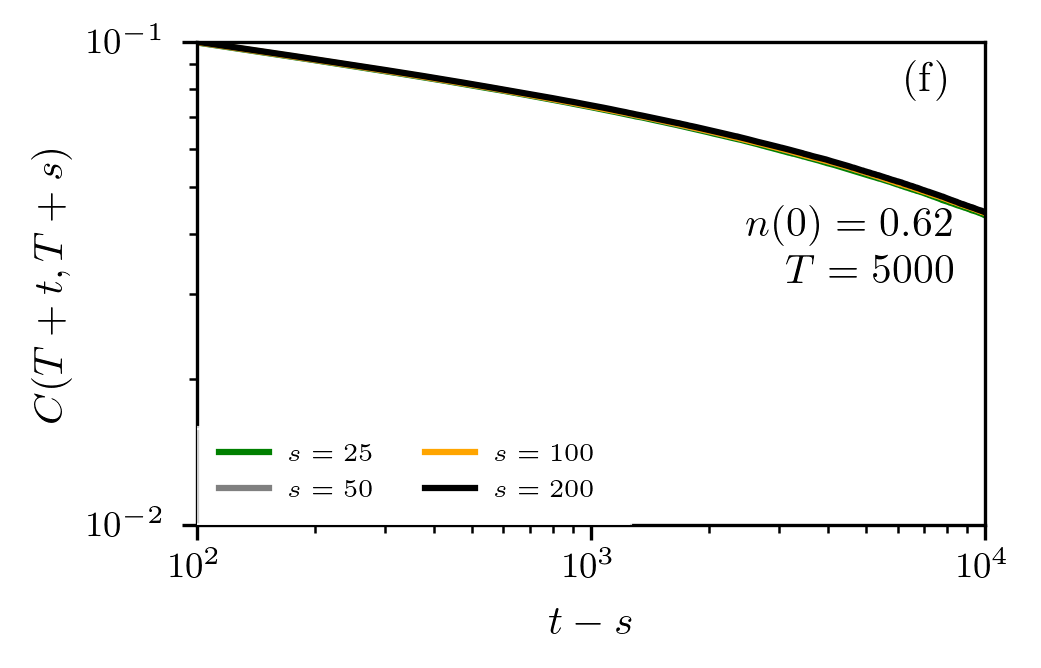}}\\
\addheight{\includegraphics[scale = \size]{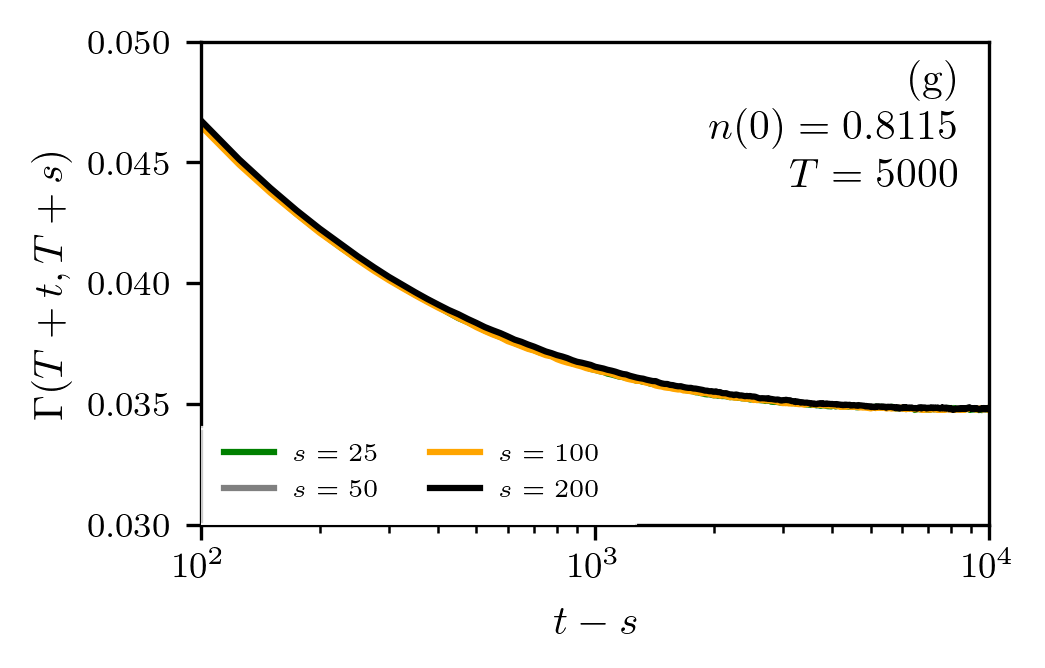}}&
\addheight{\includegraphics[scale = \size]{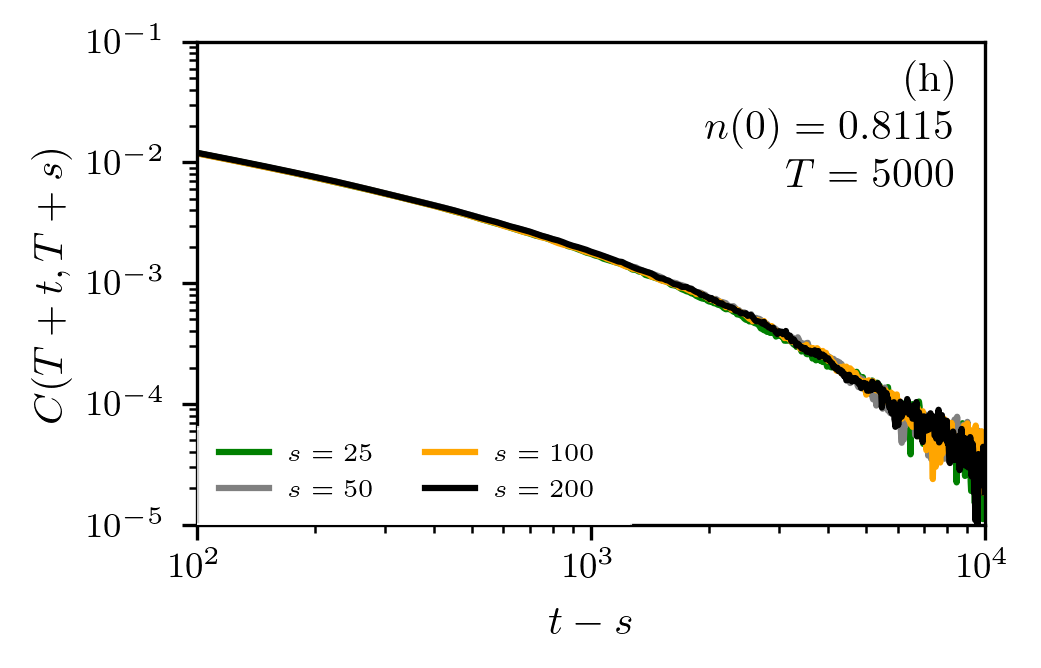}}\\
\end{tabular}
\caption{\textbf{Correlation and covariance functions for different densities of active nodes.} We show the correlation function $\Gamma(t,s)$ and covariance function $C(t,s)$ (see Eqs.~\eqref{eq:gamma} and~\eqref{eq:C}) as functions of $t-s$. In panels (a--d), we show  $\Gamma(t,s)$ and $C(t,s)$ without an initial relaxation period (i.e., $T=0$). For a comparison with longer initial relaxation times, we set $T=5000$ in panels (e--h). In panels (c--d) and (g--h), we use an initial condition of $n(0)=0.8115$ (close to the upper stationary state) and we set $n(0)=0.62=n^*(0)$ in panels (a--b) and (e--f). In our simulations, we used a square lattice with $N=512\times 512$ sites and generated $1000$ samples. As we show in panel (g), the correlation function $\Gamma$ converges towards the square of the density $n_{st}^3$ with the value $(1 - n_{st}^3)^2 \approx 0.035$.}
\label{fig:correlator1}
\end{figure*}

In Fig.~\ref{fig:correlator1}, we show $\Gamma(T+t,s)$ and $C(T+t,s)$ as functions of $t-s$ for different values of $T$ and $s$. In Fig.~\ref{fig:correlator1}(a-b), the initial density is $n(0) = 0.62 \simeq n^*(0)$ and we observe that both correlation functions are not time-translational invariant as we cannot obtain a data collapse when plotting $\Gamma(t,s)$ and $C(t,s)$ as a function of $t-s$. During the initial fast relaxation, close to the stable stationary state with density $n_{\mathrm{st}}^3 \approx 0.81$, we find that $\Gamma(t,s)=\langle n_i(t) n_i(s) \rangle$ becomes stationary since the dynamics approaches the stable stationary state exponentially fast (see Fig.~\ref{fig:correlator1}(c)). Comparing Figs.~\ref{fig:correlator1}(b) and (d) shows that the dependence of the autocovariance function $C(t,s)$ on $s$ is less pronounced for an initial density close to $ n_{\mathrm{st}}^3$ than for $n^*(0)$. However, if we let the dynamics evolve for a period $T=5000$ before determining the correlation functions, we find that $\Gamma(t,s)$ and $C(t,s)$ become time-translational invariant for both considered initial conditions (see Figs.~\ref{fig:correlator1}(e--h)). The trajectory that started close to the stable stationary state fluctuates around $n_{\mathrm{st}}^3$ and the trajectory that started at $n(0) = 0.62$ either approached one of the stable stationary states or is still located between the basins of attraction of the two stable stationary states (see Fig.~\ref{fig:inex}(c)). 

\begin{figure*}[h]
\centering
 \begin{tabular}{ccc}
 \addheight{\includegraphics[scale = \size]{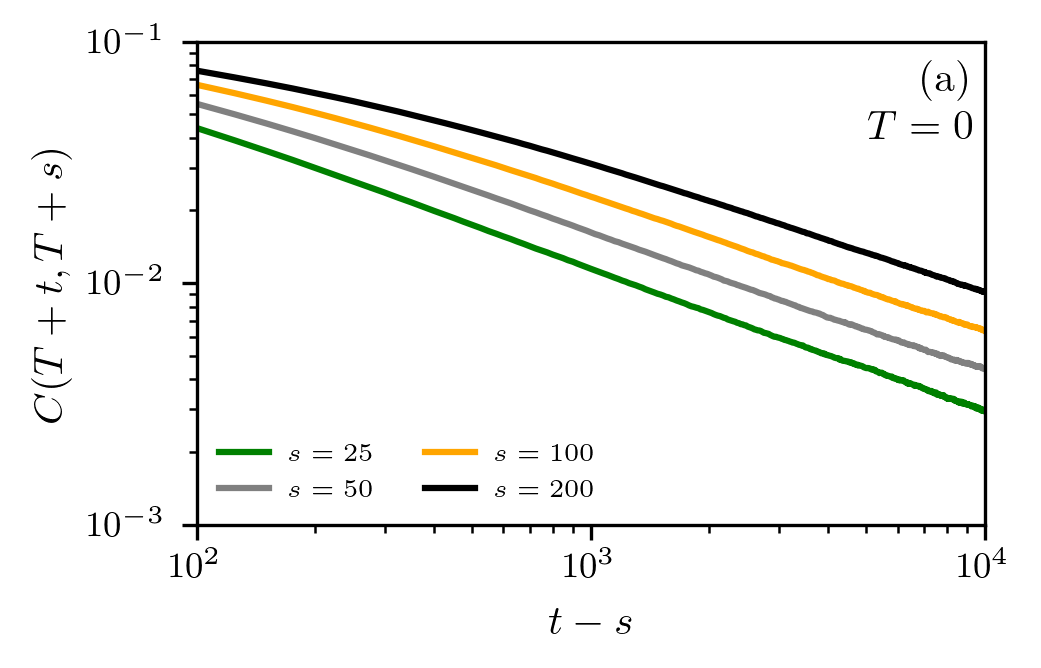}}&
 \addheight{\includegraphics[scale = \size]{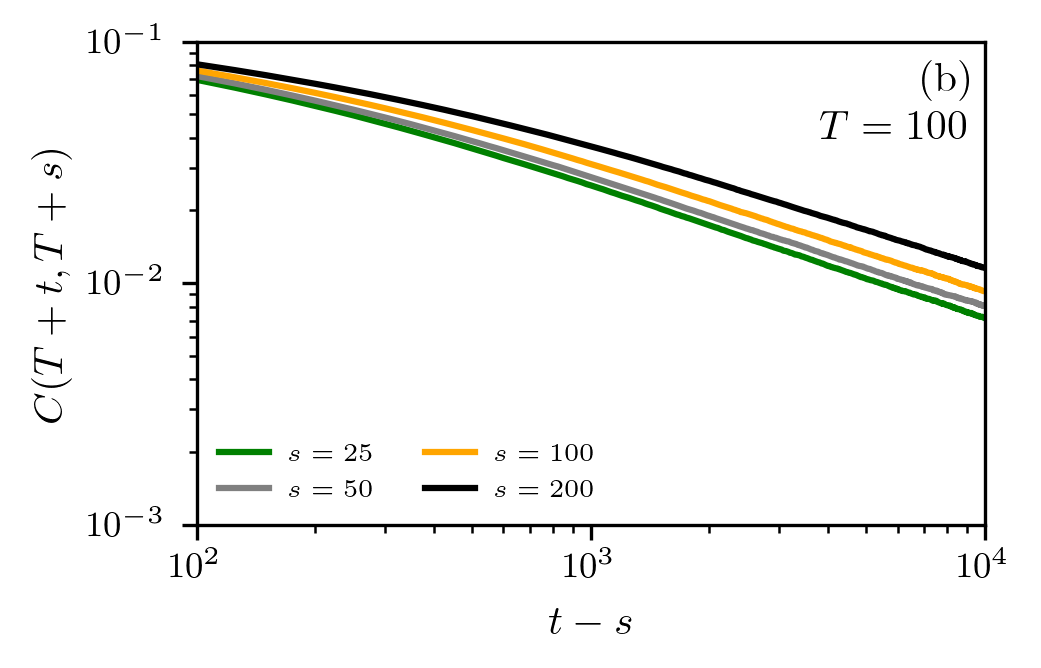}}\\
 \addheight{\includegraphics[scale = \size]{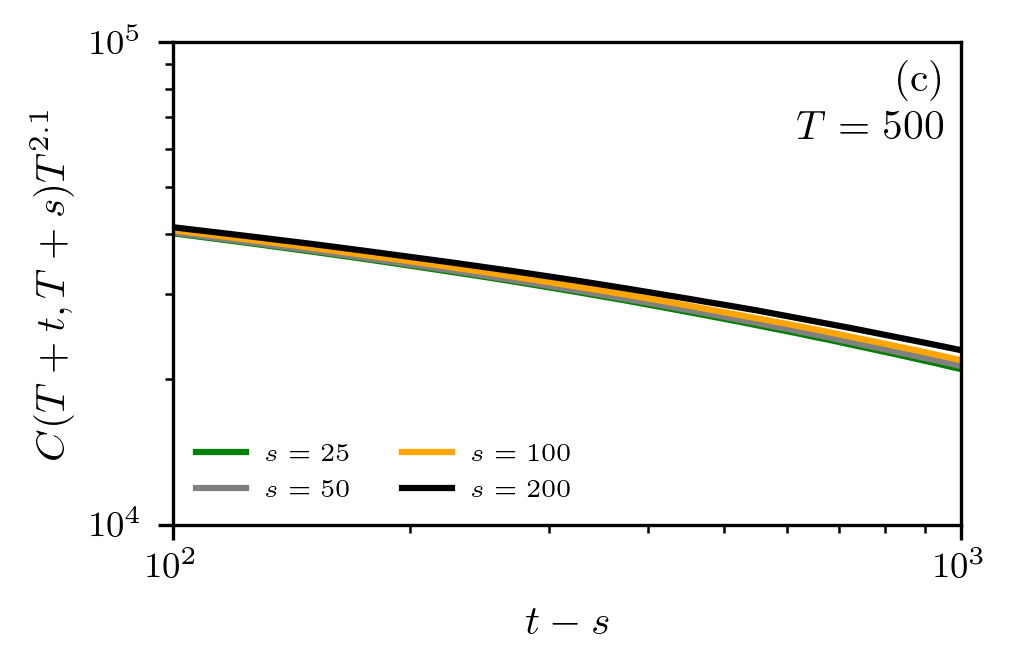}}&
 \addheight{\includegraphics[scale = \size]{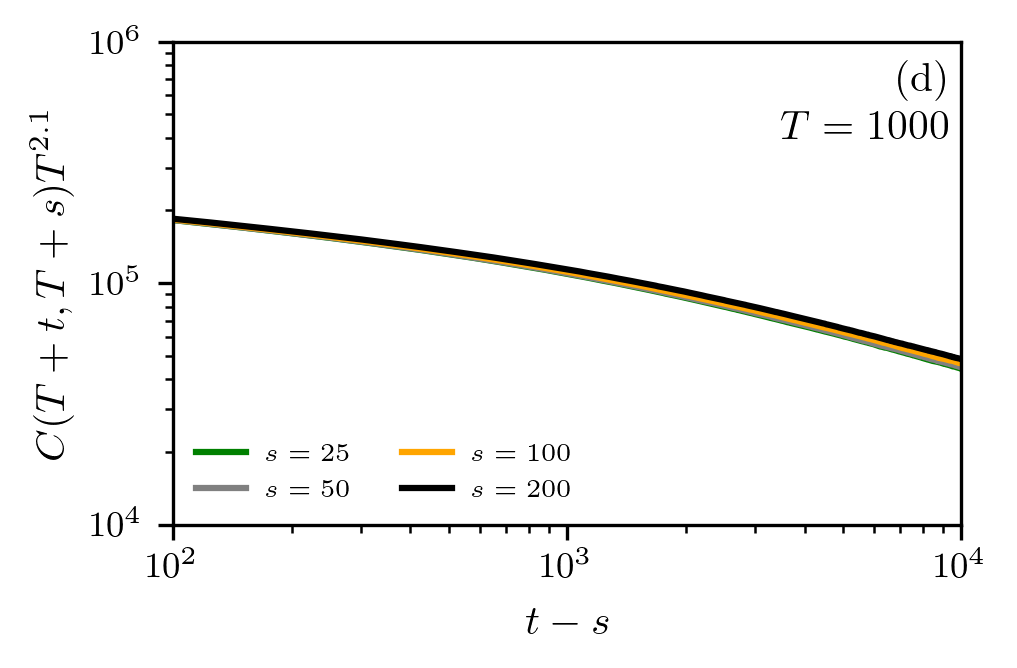}}\\
 \addheight{\includegraphics[scale = \size]{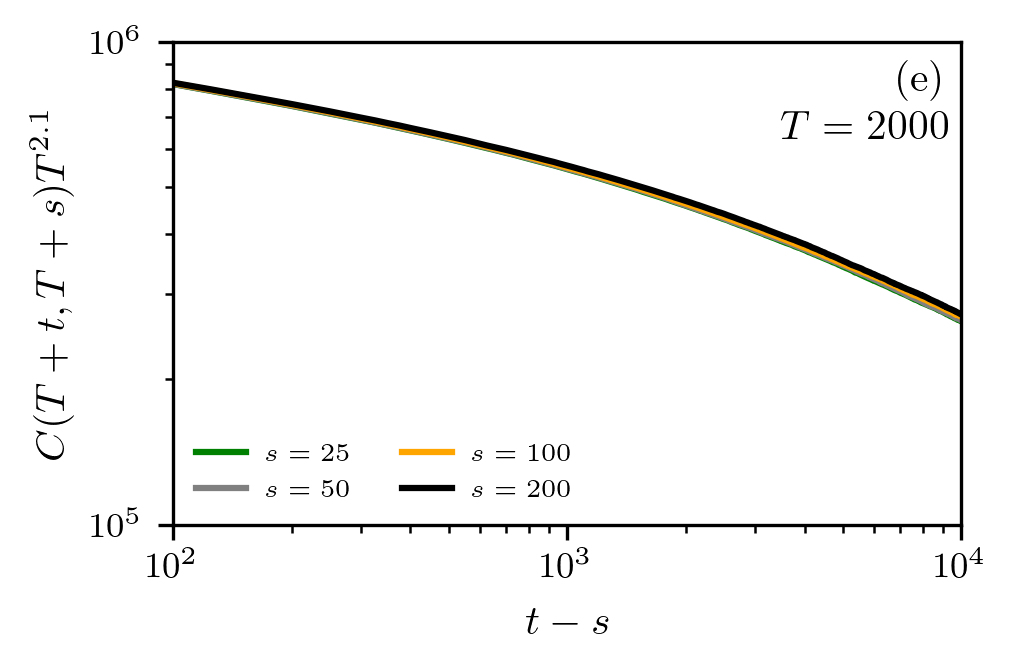}}&
 \addheight{\includegraphics[scale = \size]{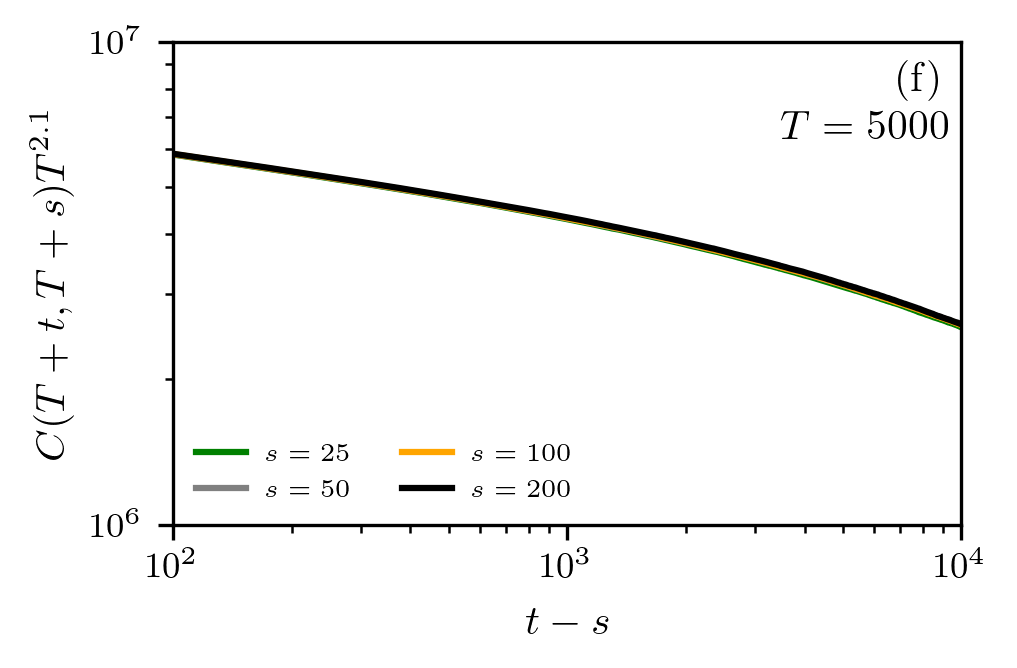}}\\
 \end{tabular}
 \caption{\textbf{Covariance functions at $n^*(0)$ for different relaxation times.} We show the covariance function $C(T+t,T+s)$ (see Eq.~\eqref{eq:C}) as a function of $t-s$. We use an initial density of 
$n(0)=0.62=n^*(0)$ and let the system relax for a period $T$ before measuring the correlation functions. The relaxation times were chosen as $T=0$ (a), $T=100$ (b), $T=500$ (c),
 $T=1000$ (d), $T=2000$ (e), and $T=5000$ (f).
 In our simulations, we use a square lattice with $N=512\times 512$ sites and generated $1000$ 
samples for each value of $T$.}
 \label{fig:correlator3}
 \end{figure*}

In Fig.~\ref{fig:correlator3}, we show $C(T+t,T+s)$ for different initial relaxation times $T\in\{0,100,500,1000,2000,5000\}$. We account for the additional $T$-dependence in Eq.~\eqref{eq:C} and obtain
\begin{equation}
C(T+t,T+s) = (T+s)^{-b} f_C( (T+t)/(T+s))\,.
\end{equation}
If $s \ll T$ and $T \ll t$, we can expand the argument of $f_C$ as follows:

\begin{equation}
\frac{T+t}{T+s} = 1 + \frac{t-s}{T} + \frac{t s}{T^2} \sim 1 + \frac{t-s}{T}\,.
\end{equation}
The resulting autocovariance function is (see Eq.~\eqref{eq:C2})
\begin{align}
\begin{split}
C(T+t,T+s) &\sim T^{-b} (1 + (t-s)/T)^{-\lambda_C/z}\,.
\end{split}
\label{eq:expansion_C}
\end{align}
We have thus shown that the function $C(T+t,T+s)$ only depends on $t-s$ for large values of $T$ (see Figs.~\ref{fig:correlator1}(e--h)). This result is in agreement with the data that we show in Fig.~\ref{fig:correlator3}. For $T=$ 500, 1000, 2000, and 5000, we observe that $C(T+t,T+s)$ is well-captured by a function of $t-s$.

\begin{figure*}[h]
\centering
\begin{tabular}{ccc}
\addheight{\includegraphics[scale = \size]{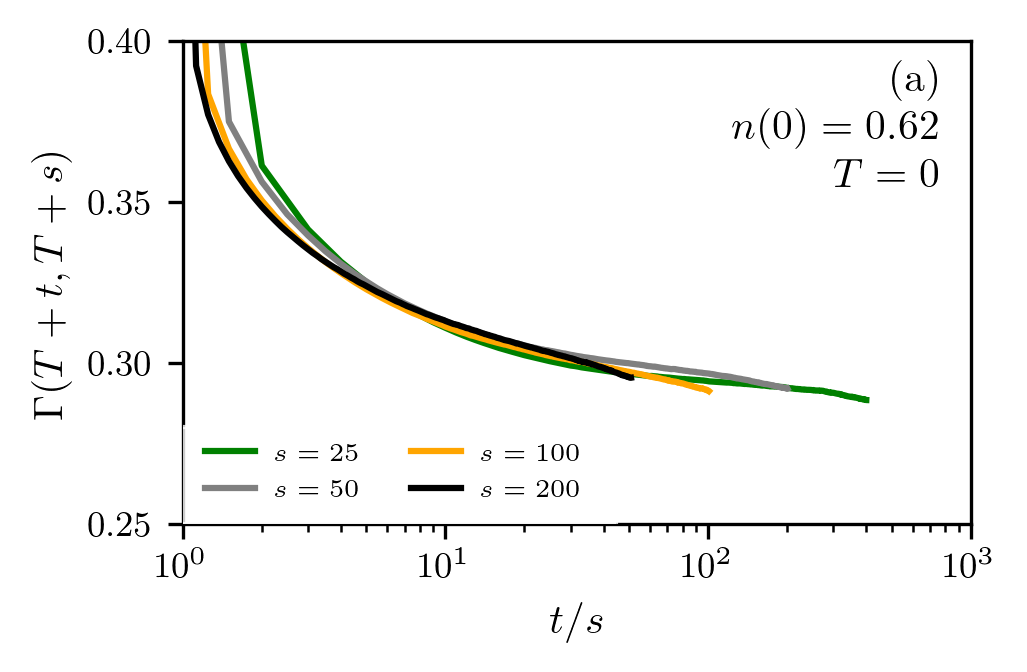}}&
\addheight{\includegraphics[scale = \size]{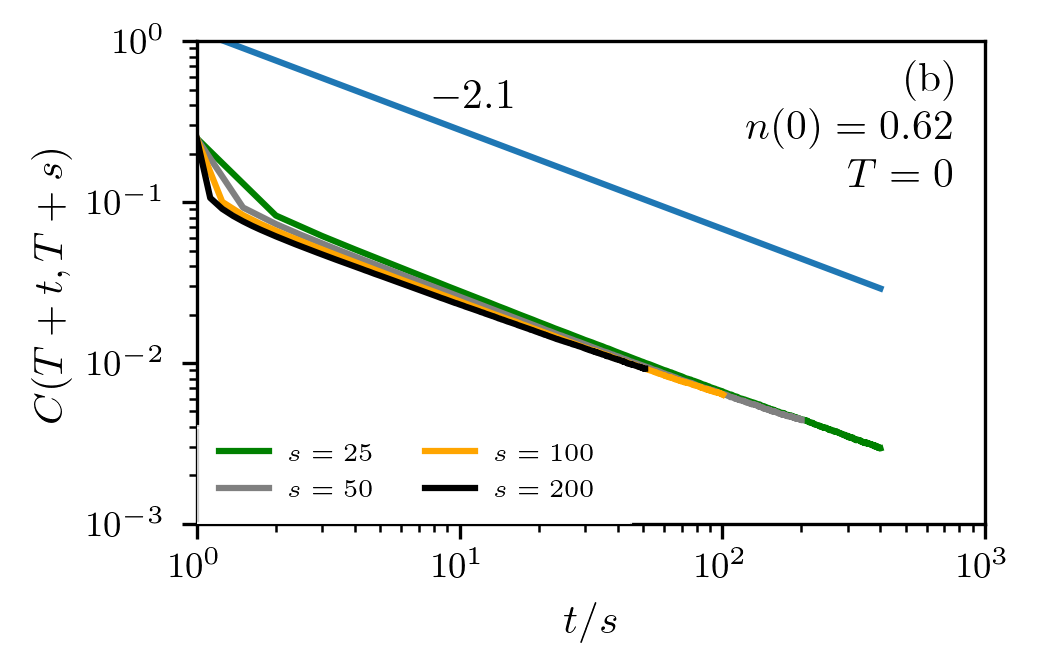}}\\
\end{tabular}
\caption{\textbf{Correlation and covariance functions at $n^*(0)$.} We show the correlation function $\Gamma(t,s)$ and $C(t,s)$ (see Eqs.~\eqref{eq:gamma} and~\eqref{eq:C}) as functions of $t/s$. In both panels (a) and (b), we use an initial density of 
$n(0)=0.62=n^*(0)$. All simulations were performed on a square lattice with $N=512\times 512$ sites and without any initial relaxation before measuring the correlation functions (i.e., $T=0$). The number of samples is $1000$.}
\label{fig:correlator2}
\end{figure*}

For the described initialization protocol, we would expect that $n(t) \propto t^{-\delta}$ with $\delta \approx 0$ for initial densities $n(0)$ close to the density $n^*(0)$. We observe in Fig.~\ref{fig:correlator2}(a) that the exponent $\lambda_\Gamma/z$ of Eq.~\eqref{eq:gamma2} satisfies the relation $\lambda_\Gamma/z = \delta \approx 0$ that was found in other systems exhibiting aging~\cite{henkel08,bottcher2018dynamical}. For the autocovariance function $C(t,s)$, we observe a data collapse when, according to Eq.~\eqref{eq:C2}, plotted as function of $s^{-b} f_{C} (t/s)$, confirming the relation $b= 2 \delta \approx 0$~\cite{henkel08,bottcher2018dynamical}. The corresponding scaling 
exponent is $\lambda_C/z= 2.1(1)$  (see Fig.~\ref{fig:correlator2}(b)). This is not too different from the estimate $\lambda_C/z=2.8(3)$ of the 2D contact process~\cite{enss2004ageing,ramasco2004ageing,bottcher2018dynamical} but not accurate enough for a quantitative comparison.
\section{Conclusions and discussion}

\label{sec:conclusion}
We have studied the relaxation properties of a general contagion model that describes simple (i.e., single contact) and complex (i.e., multiple contact) contagion phenomena. Relaxation and aging properties of simple contagions or contact processes have been analyzed in previous works~\cite{enss2004ageing,ramasco2004ageing,bottcher2018dynamical}. Here we analyzed the relaxation and aging dynamics of complex contagion phenomena within their bistable region~\cite{boettcher162,boettcher171}.

Our numerical analyses of the relaxation behavior of complex contagion dynamics show that the phase portrait and large parts of the relaxation dynamics are qualitatively captured by corresponding mean-field results~\cite{boettcher171}. However, our results also indicate that the phase space of the considered general contagion model is more complex than previously believed~\cite{majdandzic14,boettcher162,boettcher171} since it cannot be solely described by the density of failed nodes. For complex contagion dynamics on a square lattice, we have also shown that the phases in which the dynamics approach either the upper or lower stable stationary states are not linearly separable. Trajectories that cross in $(u,v)$ space at a certain time may approach different stationary states due to the influence of structural effects in the initial conditions. Aging effects can be observed for initial conditions that are close to the boundary that separates the phases in which the dynamics approach one or the other stationary state. The observed relaxation exponent $\delta$ is almost zero and the resulting aging phenomena are similar to those observed for the spherical model in an external magnetic field~\cite{paessens2003kinetic}.
Since the universality class of the 2D SIR model is thought to be the same as the one of dynamical percolation~\cite{PhysRevE.82.051921}, it would be desirable to dispose direct studies of aging in this universality class.

A further possible extension of our work would be to study the influence of different lattice structures (e.g., honeycomb, triangular, and hexagonal lattices) on the relaxation and aging properties of complex contagions. To study different network structures, one has to first numerically identify the hysteresis (or bistable) region. For the square lattice this region is extremely small and can be only identified with high resolution simulations (in $(r,p)$ space) for large systems~\cite{boettcher162}. Similar efforts may be also necessary for other lattice structures. After having identified this region, one can determine the phase portrait as in Fig.~\ref{fig:graphs1} to locate the unstable point and study the relaxation properties. Note that similar to the directed percolation universality class, one may find mean-field relaxation behavior in dimensions larger than or equal to $4$~\cite{henkel2008non}.
\acknowledgments{LB acknowledges financial support from the SNF Early Postdoc.Mobility fellowship on ``Multispecies interacting stochastic systems in biology'' and the Army Research Office (W911NF-18-1-0345). All simulations have been performed on the ETH Euler cluster.}
\bibliographystyle{apsrev4-1}
\bibliography{refs}
\end{document}